# Single-photon sources created by nature millions of years ago

D. G. Pasternak[1#], A. M. Romshin[1#], R. A. Khmelnitsky [2], G. Yu. Kriulina[3], A. A. Zhivopistsev[1], O. S. Kudryavtsev[1], A. V. Gritsienko[2], A. M. Satanin[4], I. I. Vlasov[1*]

1 – Prokhorov General Physics Institute of the Russian Academy of Sciences, 38 Vavilov str., Moscow 119991, Russia
2 – P. N. Lebedev Physical Institute of the Russian Academy of Sciences, 53 Leninskiy pr., Moscow 119991, Russia
3 – Department of Geology, Moscow State University, 1 Leninskie Gory str., Moscow 119991 Russia
4 – National Research University Higher School of Economics, 20 Myasnitskaya str., Moscow 101000, Russia

[#] – these authors contributed equally to this work
[*] – corresponding author e-mail vlasov@nsc.gpi.ru

**Abstract**

Single photons source (SPS) is a key component required by quantum communication devices. We report the finding of bright diamond-based SPS created by nature millions of years ago. It is shown that narrow (≤ 2 nm) lines observed within the 500-800 nm range in photoluminescence (PL) spectra of the surface layer of untreated Yakut diamonds rich in nitrogen and hydrogen belong to SPS. Moreover, unknown narrow-line PL observed earlier in nitrogen- and hydrogen-rich diamonds from various deposits around the world are thought to be associated with SPS. Thus, the diamond rim, which until now was sent to the dumps or, at best, used as an abrasive powder, turned out to be a valuable material suitable for use in quantum technologies.

Since the 1970s, researchers around the world have been actively involved in developing efficient single photons sources. Various methods of generating single photons have been proposed: from the simplest (faint laser irradiation) to quite complex (emission of single atoms, ions, molecules), implemented at cryogenic temperatures and high vacuum. Since the 2000s, single photons sources (SPS) have been developed based on color centers in various solid-state materials, primarily diamond [1], silicon carbide [2], and boron nitride [3]. Of these materials, diamond is of greatest interest because the SPS formed in it efficiently generate single photons even at room temperature in a narrow spectral range (1-5 nm). Recently we have found a new type of narrowband (≤1 nm) SPS in nanodiamonds (NDs) synthesized at high pressure from adamantane. A large number of such SPS emitting at room temperature in a broad spectral range (500-800 nm) were found in the NDs rich in bulk nitrogen impurity and surface hydrogen. These SPS were evidentially attributed to radiative recombination of donor–acceptor pairs formed by nitrogen donors and hydrogen-stimulated acceptors on a diamond surface. This discovery motivated us to investigate the photoluminescence (PL) properties of natural diamonds rich in nitrogen and hydrogen.

Eight diamond stones with typical sizes from 2 to 10 mm, extracted from the Yubileinaya kimberlite pipe (Yakutia, Russia), were examined. Two samples are double-side polished plates. According to their optical image, they are slightly yellowish, transparent and contains small dark inclusions (Fig. 1a). The remaining six samples were rough stones of mixed colors: cloudy-grayish, yellowish and greenish (Fig. 1b). Morphology analysis of the rough parent surfaces using scanning electron microscopy (SEM) showed the presence of a predominantly wavy relief (Fig. 2a), characteristic of a single-crystal diamond growth surface. In some surface areas, stepped square-shaped pits (resembling inverted pyramid) of micron sizes were observed (Fig. 2b). Such relief is typical for etching figures formed by partial dissolution of the diamond surface at the final stage of its formation [4]. According to the classification of Yu. L. Orlov, such diamonds from the Yakut pipe Yubileinaya can be classified as Forth Variety [5].

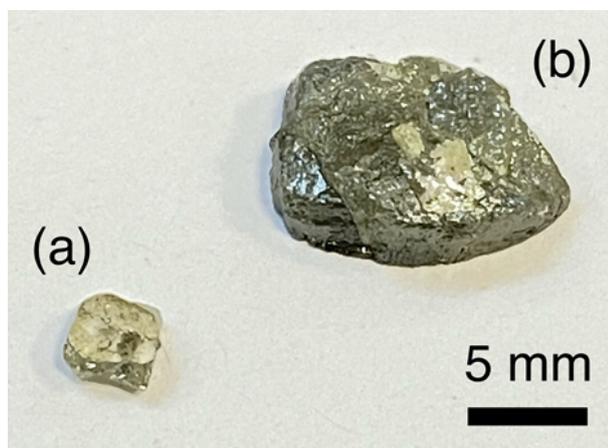

**Figure 1**. Photographic images of natural diamonds from the Yubileinaya kimberlite pipe (Yakutia, Russia). (a) double-sided polished sample (b) rough sample.

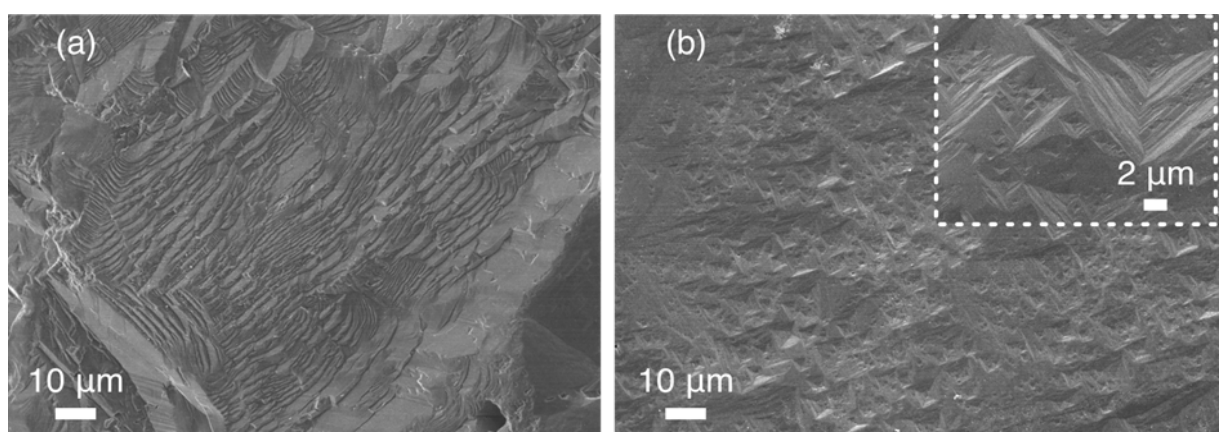

**Figure 2**. SEM images of the surface morphology obtained for rough natural diamond. (a) Region with wavy relief characteristic of a single-crystal growth surface. (b) Stepped inverted pyramids characteristic of etching figures at a late stage of diamond growth.

Assessment of nitrogen- and hydrogen-related defect concentrations in the studied samples was carried out with FIR absorption spectroscopy over the 800–4000 cm$^{-1}$ range. Absorption spectra of polished plates were recorded at room temperature using a standard Fourier-transform infrared (FTIR) spectrometer in transmission mode (see Methods). The absorption bands corresponding to the nitrogen centers of A-type dominate the spectra (Fig. 3). According to Refs [6], the A-nitrogen concentration is estimated at approximately 2000-3000 ppm. The spectra also exhibit a narrow band (3 cm$^{-1}$) at 3107 cm$^{-1}$ indicating a high hydrogen content in the diamond. Although IR absorption spectroscopy is not typically employed to quantify hydrogen concentration in diamond, it is generally accepted that a diamond is hydrogen-rich (> 500 ppm) if the intensity of the 3107 cm$^{-1}$ line is comparable to or exceeds the intrinsic two-phonon absorption at 2450 cm$^{-1}$ [7]. The presence of additional narrow lines in the 1345–1650 cm$^{-1}$ and 3000–3400 cm$^{-1}$ regions (insets in Fig. 3) is likewise associated with hydrogen-related defects within the diamond bulk [7, 8].

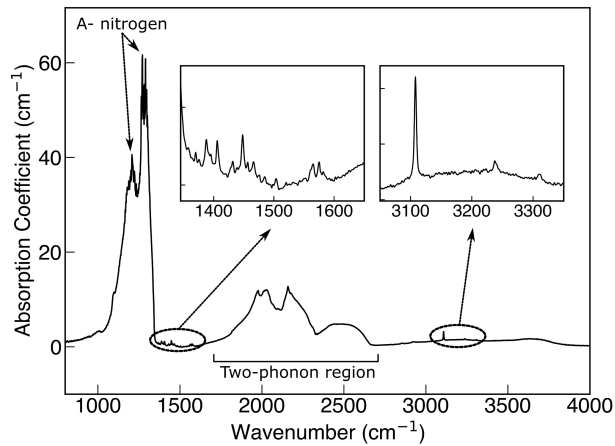

**Figure 3.** Characteristic FTIR spectrum of the polished diamond samples, recorded in transmission mode at room temperature. Absorption bands of A centers at 1282 cm$^{-1}$ and 1215 cm$^{-1}$ dominate the spectrum. (Insets) Two series of narrow absorption lines in the 1345–1650 cm$^{-1}$ and 3000–3400 cm$^{-1}$ regions attribute to H-related defects formed within the diamond bulk.

The bulk and surface PL characteristics of the investigated diamonds over the wide spectral range (500–800 nm) were studied using a confocal Raman–PL spectrometer with 473 nm laser excitation. PL spectra acquired at various surface areas of the polished and untreated samples (Fig. 4 a) are dominated by a strong emission band of two nitrogen atoms-vacancy (NVN or H3) defects centered around 520 nm., In addition, there are a number of narrow lines scattered throughout the 500–800 nm range in the spectra measured on the rough surfaces of diamond samples (Fig. 4 a,b). Similar narrow luminescent lines have been detected earlier in hydrogen- and nitrogen-rich natural diamonds mined from Argyle (Australia) and other deposits around the world [7-10]. Their origin was attributed to hydrogen impurities, but their luminescent properties have not yet been studied in detail.

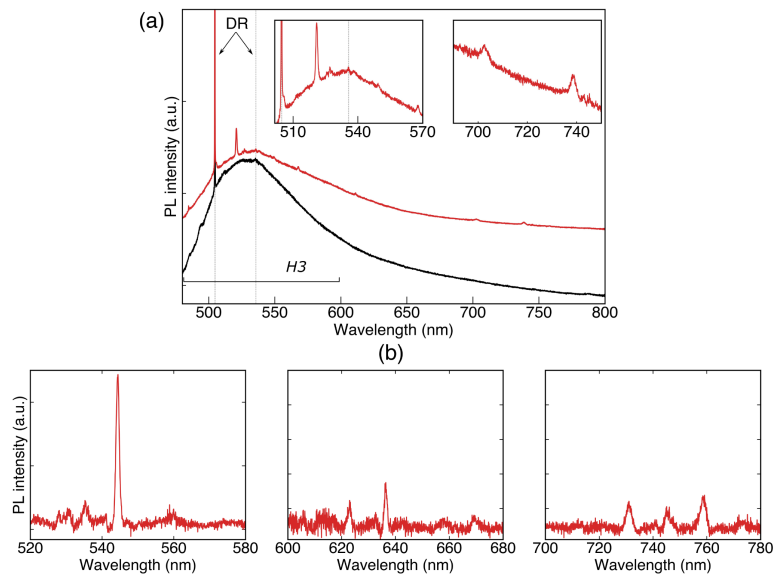

**Figure 4.** (a) Characteristic confocal PL spectra recorded under 473-nm laser excitation from the surface of the polished (black curve) and rough diamonds (red curve). Inserts: zoom-in images of spectral regions containing narrow lines. The vertical dotted lines indicate the positions of one- and two-phonon diamond Raman (DR) scattering. (b) Examples of narrow lines recorded in different spectral ranges from various surface areas of the rough diamonds.

Next the features of the spatial distribution of the narrow-line PL, as well as the spectral, power, temporal characteristics and photon statistics for a number of individual narrow lines are shown (Fig. 5). The results were obtained using a custom-built confocal luminescence spectrometer (see Methods). The PL in diamond surface layer was excited with 532 nm and 658 nm lasers to suppress background emission of nitrogen centers. PL mapping of diamond rims in various spectral ranges of signal detection indicated that narrow-line PL are predominantly found in areas with a wavy surface relief (Fig. 2a), where the density of wave "terraces" is highest. Examples of regions with the most densely packed narrow lines in the spectral interval of 727–749 nm are shown in Fig. 5a. When focusing deeper from the sample surface, the PL intensity decreases significantly and becomes undetectable at depths greater than 5 μm, indicating that the emission originates from the near-surface region. Representative PL spectra recorded from various regions of the diamond surface reveal a set of narrow lines under both 532 nm and 658 nm excitation (Fig. 5b). The widths of non-overlapping lines at their half-height vary from 0.3 nm to several nanometers. It can be seen that the lines are predominantly grouped around positions shifted from the excitation energy by a multiple of approximately 180 meV towards longer wavelengths. Under 532 nm excitation, such groups are observed around 580 nm and in the range of 620–640 nm, whereas under 658 nm excitation, the lines are grouped in the range of 720–740 nm. Additionally, the lines, comprising both Stokes and anti-Stokes components, are detected in the immediate vicinity of the excitation wavelengths.

Luminescence characteristics of individual lines were investigated under 658 nm excitation within the 700–770 nm spectral window, where the broadband background is minimal. It was determined that for the majority of lines of 0.3–2 nm width the corresponding emission demonstrates sub-Poissonian photon statistics with a second-order correlation function value of $g^{(2)}(0) < 0.5$ (Fig. 5d), i.e. they belong to single photon emitters. The saturation power for these emitters is relatively low, 1–2 mW (Fig. 5e), and the excited-state lifetimes vary in the range from 1 to 1.8 ns (Fig. 5f). The weak (<10%) contribution of a longer decay time to the luminescent kinetics is attributed to the H3 background. The maximum count rate recorded for some emitters reaches ~$10^6$ counts/s (Fig. 5e). Emission intensity from certain emitters remains stable over time, whereas most exhibit intermittent blinking without complete bleaching over the entire observation period (~100 min).

Narrow luminescent lines with similar characteristics were recently found in nitrogen- and hydrogen-rich NDs synthesized from hydrocarbons at high pressure and high temperature (HPHT technique) [11]. We explained the origin of those novel lines by radiative recombination of donor–acceptor pair (DAP) formed by donor-type nitrogen centers and acceptor states induced by hydrogen at a diamond surface. IR spectra of the natural diamonds under study reveal a substantial hydrogen content in their volume. Therefore, it is reasonable to assume that their surfaces are at least partially terminated with hydrogen. The significantly higher (by about 2 orders of magnitude) concentration of hydrogen on a surface of hydrogen-rich natural diamond compared to the volume, found in the work of J.P.F. Sellschop et al. [12], supports our assumption. Vibrational modes of surface $CH_x$ for NDs smaller than 50 nm are reliably detected by IR absorption [13]. The minute surface-to-volume ratio of the studied diamonds excludes the observation of such modes in IR spectra. In this work we confirm the presence of surface hydrogen by annealing of natural diamond in the air [14]. As we have shown earlier, removal of surface hydrogen leads to the disappearance of narrow luminescent lines [11]. A similar effect was observed at 750 °C annealing of one of the rough diamond samples (Fig. 5h).

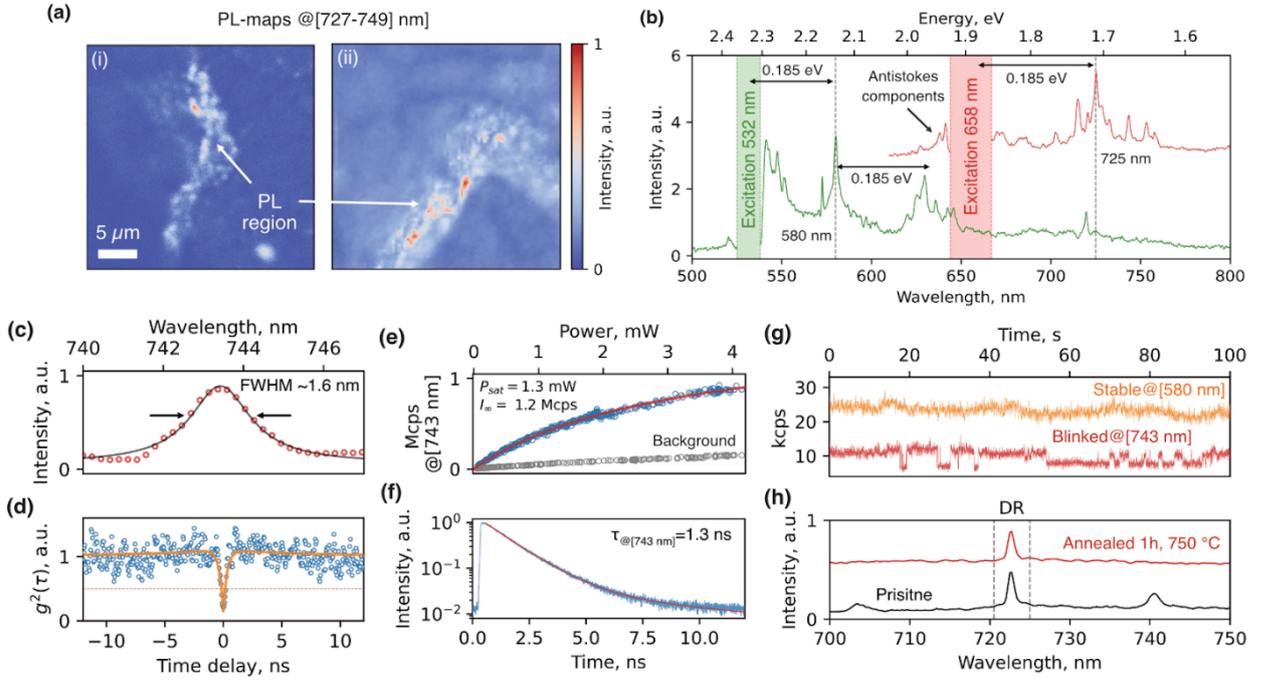

**Figure 5.** (a)Two characteristic PL maps acquired in the 727–749 nm spectral range under 658 nm laser excitation (1 mW at the sample). (b) Representative PL spectra recorded under 532 nm (green curve) and 658 nm (red curve) excitation, both at 1 mW power. (c) Zoom-in PL spectrum of individual narrow line at 743 nm (@743 nm) and fitting with Lorentz profile (black solid). The linewidth (FWHM) is 1.6 nm. (d) Second order correlation function measured for @743 nm. (e) Saturation curve for @743 nm, (f) Luminescent decay curve corresponds to almost pure monoexponential decay with time constant 1.27 ns. (g) Time traces of photon count rate from two separate lines: stable @580 nm under 532 nm (1 mW) excitation (orange) and blinking @743 nm under 660 nm (1 mW) excitation (red). (h) PL spectra measured with 658 nm excitation before and after annealing of the diamond sample at 750 °C during 1 hour. Narrow lines at 703 nm and 741 nm disappear while the diamond Raman (DR) line remains after annealing.

We believe that the observed feature in the spectral distribution of the narrow lines: their grouping with a step multiple of 180 meV (Fig. 5b), can be associated with the energies of bending (around 180 meV) and stretching (around 360 meV) vibrations of surface CHx groups and their overtones. The interaction of these vibration modes with hole states induced by surface hydrogen results in vibronic sublevels in electron DAP transitions. Donor-acceptor pairs are formed when hole states are filled with electrons from higher electron levels of donor nitrogen centers located near the surface. It is further assumed: when diamond is irradiated with 473-nm laser light, non-resonant excitation and subsequent radiative recombination of these donor-acceptor pairs occurs in a wide spectral range of 500-800 nm (Fig. 4). The width of the range is determined by the different distances of the nitrogen centers from the diamond surface, as well as the energies of the donor and acceptor levels [11]. When moving to the excitation energies lying within this range, the most effectively excited DAPs are those with energies of purely electronic or vibronic (multiples of 180 meV) transitions close to the excitation energy (Fig. 5b). Interestingly, the work of Nelyubov et al. [15] was reported the detection of a few narrow PL lines from the "enigmatic" color centers in hydrogen-terminated HPHT NDs, which were also accompanied by weak vibronic replicas with a step multiple of approximately 180 meV. We believe that the origin of those lines is associated with the same DAP emitters as in [11].

We have previously shown that the orientation of the emitting DAP dipoles is predominantly perpendicular to the surface [11], and it allows them to interact over long distances.

This property paves the way for engineering 2D arrays of quantum defects that can be exploited for scalable quantum sensing and quantum communication applications [16].

In conclusion, we have found numerous SPSs in the spectral range of 500-800 nm on the rough surface of diamond crystals mined from the Yubileinaya kimberlite pipe (Yakutia, Russia). These diamonds are shown to be characterized by high concentrations of nitrogen and hydrogen, suggesting the presence of similar SPSs in nitrogen- and hydrogen-rich diamonds from other deposits around the world. These SPSs are assumed to be donor-acceptor pairs associated with nitrogen and hydrogen on the diamond surface. In their spectral characteristics, DAP-related SPS are superior to previously known SPS formed by various impurity defects in the diamond volume. They have the narrowest emission line width and the highest photon emission rate at room temperature among known SPSs in diamond. Previously the SPS with similar characteristics were produced in nitrogen-rich and hydrogen-terminated synthetic NDs. Thus, it turned out that nature was millions of years ahead of us in creating SPS based on diamonds with a high nitrogen and hydrogen content.

**Methods**

Scanning Electron Microscopy

The diamond surface morphology was analyzed using a JEOL JSM-7001F Scanning Electron Microscope (SEM). To render the intrinsically insulating diamond surfaces conductive, a "soft" surface graphitization was applied: selected samples were annealed in a high-vacuum (~$10^{-4}$ Pa) furnace at 1200 °C for 1 hour, promoting the formation of a thin (5–7 nm) conductive layer [17].

Fourier Transform Infrared absorption spectroscopy (FTIR)

IR absorption spectra of the double-polished, flat–parallel diamond plates were acquired at room temperature using a PerkinElmer Spectrum 100 FTIR spectrometer in transmission mode. Measurements were performed over the 800–4000 cm$^{-1}$ range with a spectral resolution of 2 cm$^{-1}$, an accumulation time of 10 min. Prior to FTIR analysis, all samples underwent a mild thermal cleaning step to remove residual organic contaminants: they were air annealed at 300 °C for 1 h with a Linkam TS1500 temperature controlled stage.

The PL analysis

The bulk and surface PL characteristics of the diamond samples over the wide spectral range (500–800 nm) were studied at room temperature using a LabRam HR800 (Horiba) confocal spectrometer. PL was excited with diode lasers at 473 nm and collected in a back-scattering geometry using an Olympus microscope objective (×50 magnification, numerical aperture NA = 0.55). The power of the laser irradiation incident on a diamond surface was 1 mW. To demonstrate the relation of narrow lines with surface terminating groups, one of the rough diamond sample underwent oxidative annealing in air at 750 °C for 1 hour (Linkam TS1500).

The PL properties of the diamond samples were investigated using a custom-built optical microscope. Precise sample positioning relative to the laser focus was achieved using a three-axis piezoelectric stage (piezosystem jena) within the field of view of the CMOS camera. Two laser sources, operating at wavelengths of 532 nm and 658 nm were employed for PL excitation. The PL excitation and collection were performed using a Mitutoyo ×100 objective (NA = 0.7). The collected light was directed through a system of notch/bandpass filters and directed to the detection path. Emission spectra were recorded using a commercially available spectrometer (Optosky

ATP5200), equipped with a fiber-coupled input and a 50 μm entrance slit, with an acquisition ranging from 200 ms to 30 s. For PL mapping, avalanche photodiodes (APDs, Excelitas SPCM-AQRH-14-FC) were employed in combination with band-pass filters. Since narrow lines typically superimposed on a broad PL background, in most measurements we employed a SolarsLab M-266 monochromator in order to enhance signal-to-noise ratio by selecting the narrow spectral band and to perform spectral scanning. The spectral bandwidth was adjusted through the exit slit and varied from 0.6 nm to 2 nm, while the confocality of the optical setup was ensured by narrowing the monochromator's entrance slit. This configuration was employed for recording PL spectra and maps, as well as time stability. To confirm the single-photon nature of the individual narrow lines, a Hanbury-Brown-Twiss (HBT) interferometer with two APDs (Excelitas SPCM-AQRH-14-FC) was used. Luminescence decay rates of the narrow-line emission were measured using a Time-Correlated Single Photon Counting (TCSPC) module (PicoQuant MicroTime 200) with an Olympus ×50 objective (NA=0.55) and pulsed laser source 675 nm.